\documentclass{article}
\usepackage{spconf,amsmath,graphicx}
\usepackage[colorlinks]{hyperref}
\usepackage{xcolor}
\usepackage{amssymb}
\usepackage{booktabs}

\newcommand{\etal}[0]{\textit{et al.}}
\usepackage{multirow}
\usepackage{adjustbox}
\usepackage{array}
\usepackage[utf8]{inputenc}
\ninept
\title{SpatialCodec: Neural Spatial Speech Coding}
\name{Zhongweiyang Xu$^{\star}$\thanks{$^{\star}$This work was done while Z. Xu, H. Wang, and M. Yang were interns at Tencent AI Lab, Bellevue, USA.} \qquad Yong Xu$^{\dagger}$ \qquad Vinay Kothapally$^{\dagger}$ \qquad Heming Wang$^{\sharp}$ \qquad Muqiao Yang$^{\ddagger}$ \qquad Dong Yu$^{\dagger}$}

\name{
\begin{tabular}{c}
     Zhongweiyang Xu$^{\star}$\thanks{$^{\star}$This work was done while Z. Xu, H. Wang, and M. Yang were interns at Tencent AI Lab, Bellevue, USA.}, Yong Xu$^{\dagger}$, Vinay Kothapally $^{\dagger}$, Heming Wang $^{\sharp}$, Muqiao Yang $^{\ddagger}$, Dong Yu $^{\dagger}$
\end{tabular}
}
\address{
$^{\dagger}$Tencent AI Lab, $^{\star}$ University of Illinois at Urbana-Champaign, \\
$^{\sharp}$ The Ohio State University, $^{\ddagger}$ Carnegie Mellon University
}

\begin{document}
%
\maketitle
\begin{abstract}

In this work, we address the challenge of encoding speech captured by a microphone array using deep learning techniques with the aim of preserving and accurately reconstructing crucial spatial cues embedded in multi-channel recordings. We propose a neural spatial audio coding framework that achieves a high compression ratio, leveraging single-channel neural sub-band codec and SpatialCodec. Our approach encompasses two phases: (i) a neural sub-band codec is designed to encode the reference channel with low bit rates, and (ii), a SpatialCodec captures relative spatial information for accurate multi-channel reconstruction at the decoder end. In addition, we also propose novel evaluation metrics to assess the spatial cue preservation: (i) spatial similarity, which calculates cosine similarity on a spatially intuitive beamspace, and (ii), beamformed audio quality. Our system shows superior spatial performance compared with high bitrate baselines and black-box neural architecture. Demos are available at \href{https://xzwy.github.io/SpatialCodecDemo/}{https://xzwy.github.io/SpatialCodecDemo}.
Codes and models are available at \href{https://github.com/XZWY/SpatialCodec}{https://github.com/XZWY/SpatialCodec}.



\end{abstract}
\vspace{-4pt}
\begin{keywords}
\vspace{-4pt}
microphone array, audio codec, spatial audio
\end{keywords}

\vspace{-8pt}
\section{Introduction}
\vspace{-7pt}
\label{sec:intro}

Audio and speech codec aims at compressing the signals into low bitrate codes for efficient storage or network streaming applications. Traditionally, these coding schemes usually take advantage of some signal models and psycho-acoustics. Speech codecs like CELP~\cite{CELP}, SPEEX~\cite{speex}, OPUS~\cite{opus}, and EVS~\cite{EVS} all use linear predictive modeling for signal analysis. Likewise, audio or music codecs like MP3~\cite{mp3} and OPUS~\cite{opus}, apply classic perceptual coding technique~\cite{perceptual_coding} inspired by the perceptual masking effect of human hearing. In addition, conventional quantization and entropy coding methods~\cite{quantization} are applied for discretization and efficient coding respectively in these classic codecs. However, the performance of these conventional methods suffers with very low bit rates, e.g., at 6 kps.

While traditional codec(s) struggle to achieve high-quality, perceptually accurate reconstruction at extremely low bitrates, neural codecs are capable of overcoming this limitation. SoundStream~\cite{soundstream} uses time-domain CNNs as encoding and decoding blocks with a residual vector quantization (RVQ) to compress the intermediate latent representations. TF-Codec~\cite{TF-Codec} uses a temporal linear predictive coding technique to further remove temporal redundancies. More recently, Encodec~\cite{encodec} has been proposed which employs transformer-based network to model the code distribution, an approach originally intended for arithmetic coding, in order to achieve improved compression. AudioDec~\cite{audiodec} further proposes a two-stage training scheme such that the encoder and decoder can be more flexible and easily switched for different applications. HiFi-Codec~\cite{hificodec} substitutes the RVQ block with grouped residual vector quantization (GRVQ) and achieves better performance. These aforementioned neural approaches are primarily derived or adapted from VQ-VAE~\cite{vq-vae} and GAN-based Vocoders~\cite{melgan, hifigan}. Thus, the encoded information from these codec(s) is also treated as learned representations for generative tasks~\cite{speech_resynthsis, audio_craft, valle}.

Besides single-channel codecs, spatial audio codec aims to compress multi-channel audio while preserving the spatial information~\cite{yang2005high}. Such multi-channel codecs are commonly designed for playback systems, or multi-speaker systems, e.g., stereo coding~\cite{stereo}, MPEG-3D Audio~\cite{mpeg_3daudio}, MPEG-Surround~\cite{mpeg_surround}. Typically, a spatial audio codec adheres to a pipeline consisting of the following steps: (i) downmix the multi-channel audio into mono or stereo, and code with a traditional audio codec, (ii) some sub-band spatial parameters are extracted from the multi-channel audio and coded channel-wise and band-wise, (iii) The decoder then resynthesizes the multi-channel audio from the previous two components. Since these codecs are designed for specific playback systems, they do not consider microphone array spatial recording systems. Bao~\etal~\cite{mimo_speex} applies this scheme to microphone array recorded speech but without any reverberation (only 
the direct path exists). These methods do not use any neural networks and also do not fully exploit inter-channel or inter-band correlations. This means for a decent reconstruction, the system needs to have reasonably large bands coded separately for each channel, resulting in high coding bit rates.
\begin{figure*}[ht]
\vspace{-1cm}
    \centering
          \includegraphics[scale=0.5]{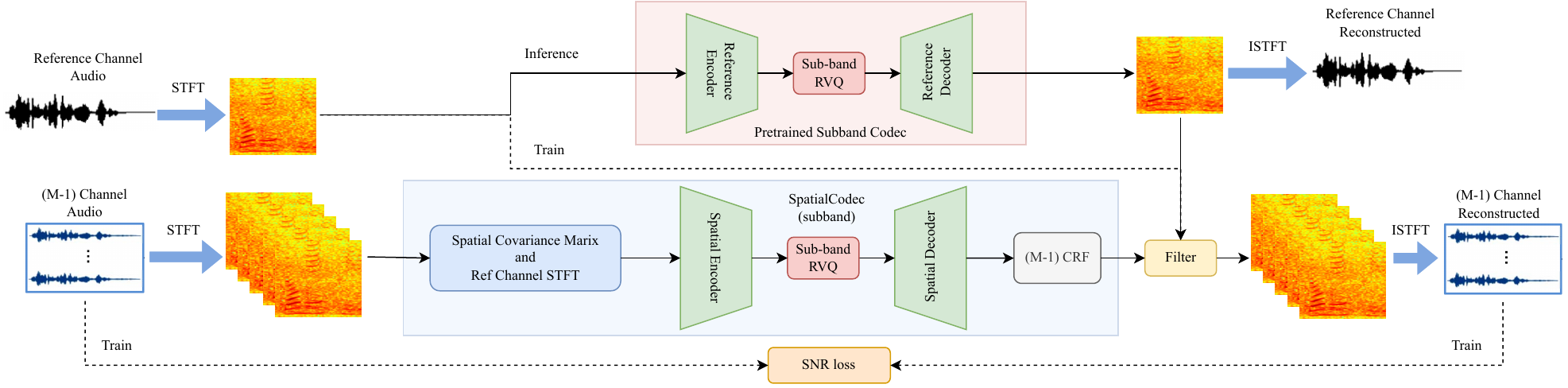}
          \vspace{-0.5cm}
    \caption{An Overview of the proposed SpatialCodec framework.}
    \label{fig:Overall}
    \vspace{-0.6cm}
\end{figure*}

Our SpatialCodec aims to address the high coding rate challenge with neural networks. Similar to the conventional methods, our design also has two branches: the first branch codes the reference channel audio, while the second branch codes the spatial information. Then on the decoder side, the first branch's decoder outputs the reconstructed reference channel. Then the second branch's decoder's output and the reconstructed reference channel are used jointly to synthesize all non-reference channels. 
We train and test our proposed codec on a synthesized multi-channel spatially rich reverberant dataset with speech from a single speaker. 


We also propose several novel metrics to evaluate spatial cue preservation. One is spatial similarity, which calculates the cosine similarity between the estimated and ground-truth spatial features. Spatial features are designed by beamforming towards a few fixed directions. We believe this metric is a more intuitive metric because the spatial features are directly related to real-world directions. We also propose to use beamforming performance as a metric to validate our SpatialCodec's ability to preserve both the spectral quality and the main directivity. 
Our proposed SpatialCodec with 12 kbps of bitrate performs significantly better than 96 kbps (8 channels x 12 kbps/channel) OPUS and other channel-independent neural codecs. We also designed one black-box model for comparison. 


\vspace{-10pt}
\section{Problem Formulation}
\vspace{-9pt}


We consider the challenge of compressing spatial audio recordings from a $M$-channel microphone array in a reverberant environment. Let $s(t)$ and $h_i(t)$ represent the clean speech from the speaker and room impulse response (RIR) from the speaker to the $i$-th microphone. The signals captured by an $M$-channel microphone array, $\mathbf{x}(t)$ (termed as “mixture") at time `$t$' is defined as:

\vspace{-1.1em}
\begin{equation}
   \mathbf{x}(t) = \big[h_1(t), ...., h_M(t)\big] \ast s(t)
\end{equation}
\vspace{-1.2em}

where `$\ast$' denotes the convolution operation, and $\mathbf{x}(t) = \big[x_1(t), ...., x_M(t)\big]$ includes speech captured from the speaker's direct path as well as early and late reflections using the $M$-channel array. The goal of the proposed model is to compress $\mathbf{x}$ to a low-bitrate representation `$\mathbf{C}$' such that the reconstructed multi-channel audio $\hat{\mathbf{x}}$ at the decoder preserves all spatial cues. Our proposed model consists of an encoder ($\Psi_\mathrm{Enc}$), a quantizer ($\Psi_\mathrm{Quant}$), and a decoder ($\Psi_\mathrm{Dec}$) which are jointly optimized such that $\hat{\mathbf{x}}$ is approximating $\mathbf{x}$ from both spectral (perceptual quality) and spatial (direct path, early, and late reflections) perspectives.

\vspace{-1.85em}
\begin{equation}
   \hat{\mathbf{x}} = \Psi_\mathrm{Dec}\big(\mathbf{C}\big); \quad \text{where} \quad \mathbf{C} = \Psi_\mathrm{Quant}\Big(\Psi_\mathrm{Enc}\big(\mathbf{x}\big)\Big)
\end{equation}

\vspace{-18pt}
\section{Method}
\vspace{-9pt}
\label{sec:Method}
The overall architecture of the proposed SpatialCodec is depicted in Fig.\ref{fig:Overall} which comprises two main branches: (i) a single-channel sub-band codec pre-trained to code the reference channel of the microphone array, and (ii) a SpatialCodec that codes spatial information to reconstruct multi-channel audio signals.

\vspace{-10pt}
\subsection{Single-Channel Sub-band Codec (First Branch)}
\vspace{-5pt}
The reference channel sub-band codec is a neural frequency domain sub-band codec. The reason we are designing this instead of using existing time-domain codecs is that the structure aligns with the SpatialCodec in the frequency domain, which will be discussed in~\ref{spatialcodec}. The input $x_{\text{ref}}\in \mathbb{R}^{2\times T \times F}$ is the STFT of the reference channel audio, where $2$ corresponds to real and imaginary components. Then the whole encoder-decoder architectures are 2D-CNNs with residual blocks treating real-imaginary as the channel dimension. The architecture is similar to HiFi-Codec \cite{hificodec} except 1-D CNN becomes 2-D CNN and downsampling time becomes downsampling frequency.

Encoder and Decoder all have six convolutional layers, each followed by a residual unit. For the encoder, The six layers' kernel and stride for the time dimension are always 3 and 1, respectively. The kernels and strides for the frequency dimension are $[5,3,3,3,3,4]$ and $[2,2,2,2,1]$, respectively. The output channel dimensions for all layers are $[16, 32, 64, 128, 128, 256]$. We use 640-point FFT, which means the encoder would compress the frequency dimension from 321 to 6 convolutional sub-bands. Then we code these 6 sub-bands independently using residual vector quantization. The decoder is just the opposite transpose convolution version of the encoder.

Each residual unit contains two residual blocks. Each block contains three 2-D time-dilated CNN layers with skip connections. The first block's three layers' kernel and dilation sizes are $[(3,3), (3,5), (3,5)]$ and $[(1,1), (3,1), (5,1)]$, respectively, in order of (time, freq). The second block's corresponding configurations are $[(7,3), (7,5), (7,5)]$ and $[(1,1), (3,1), (5,1)]$. Details can be checked in the source code.

\vspace{-9pt}
\subsection{SpatialCodec (Second Branch)}
\vspace{-3pt}
\label{spatialcodec}

The SpatialCodec has the same structure as the reference channel codec, except input, output, and channel dimensions of the six convolutional layers.
The input of SpatialCodec is the reference channel STFT and spatial covariance matrix concatenated in the channel dimension. Given $M$-channel STFT $\mathbf{X}(t, f)\in \mathbb{C}^{M \times 1}$, the spatial covariance matrix $\mathbf{\Phi}(t, f)\in \mathbb{C}^{M \times M}$ is defined to be:
\vspace{-7pt}
\begin{equation}
   \mathbf{\Phi}(t, f) = \mathbf{X}(t, f)\mathbf{X}(t, f)^\text{H}
\vspace{-4pt}
\end{equation}

Then the real and imaginary parts of $\mathbf{\Phi}(t, f)$ are concatenated with the real and imaginary part of the reference channel STFT, which gives a $2(M^2{+}1)$  dimensional real feature for each time-frequency bin. The feature is then treated as the channel dimension when fed into the SpatialCodec. The output channel dimensions for all the layers in the encoder are $[128, 128, 128, 128, 256, 256]$. Otherwise, the encoder and quantizer are the same as the reference channel codec. For the spatial decoder, the output is $M{-}1$ complex ratio filters (CRFs)~\cite{deepfilter} $W_m(t, f) \in \mathbb{C}^{2L{+}1, 2K{+}1}, m\in[1, ..., M{-}1] $ for all the non-reference channels. CRFs encode the spatial relative transfer functions. Assume the output of the reference channel STFT is $\hat{X}_{ref}\in\mathbb{C}^{T\times F}$, then all non-reference channels for $m\in[1, ..., M{-}1]$ are:

\vspace{-15pt}
\begin{equation}
\label{eq:inference}
   \hat{X}^{\text{non\_ref}}_{m}(t, f) = \sum_{l=-L}^{L}\sum_{k=-K}^{K}W_m(t,f,l,k) \hat{X}_\text{ref}(t{+}l, f{+}k)
   \vspace{-3pt}
\end{equation}

Thus the last layer of the spatial decoder's output channel dimension is $2\times(2L+1) \times (2K+1) \times (M-1)$, where the first 2 means real and imaginary.

\vspace{-8pt}
\subsection{Training and Loss}
\vspace{-4pt}
The training loss of the first branch mostly follows the HiFi-Codec's \cite{hificodec} strategy, using reconstruction loss, adversarial loss, and codebook learning loss. We also add an additional time-domain SNR loss with weighting $\lambda$=5. The SNR loss is defined as:

\vspace{-5pt}
\begin{equation}
\label{eq:snr_loss}
   L_\text{SNR}(x,\hat{x}) \triangleq - 10\text{log}_{10}(\frac{||x||^2}{||x-\hat{x}||^2})
\end{equation}

The second branch is trained separately with respect to the first branch. During training, the complex ratio filter (CRF)~\cite{deepfilter} is applied to the original reference channel audio instead of the reconstructed reference channel from the first branch. The reason is that even after pretraining, the first branch can only output reconstructed speech that's perceptually equivalent to the original speech. The underlying spectrogram or waveform does not have an exact match. This means if we apply the CRF to the reconstructed speech, the original non-reference channel audio cannot be used as learning targets, because of the mismatching problem of the first branch. Thus during training, in contrast to Eq.~\ref{eq:inference}, we have
\vspace{-7pt}
\begin{equation}
\label{eq:train}
   \hat{X}^{\text{non\_ref}}_{m}(t, f) = \sum_{l=-L}^{L}\sum_{k=-K}^{K}W_m(t,f,l,k) X_\text{ref}(t{+}l, f{+}k)
   \vspace{-6pt}
\end{equation}
where $\hat{X}_{\text{ref}}$ is substituted as the original reference channel signal $X_{\text{ref}}$. Then we simply use the time domain SNR loss averaged over all non-reference channels:


\vspace{-14pt}
\begin{align}\label{eq:overall_snr_loss}
L_{\text{all}}= \frac{1}{M-1}\sum_{m=1}^{M{-}1} L_\text{SNR}(\text{ISTFT}(X^{\text{non\_ref}}_m), \text{ISTFT}(\hat{X}^{\text{non\_ref}}_m))
\end{align}

\vspace{-20pt}
\subsection{Inference}
\vspace{-5pt}
In inference, we still get $\hat{X}^{\text{non\_ref}}_{m}(t, f)$ using Eq.\ref{eq:inference}. The first branch's sub-band codec reconstructs the reference channel audio. Then the second branch's SpatialCodec reconstructs $M{-}1$ complex ratio filters, which are applied to the reconstructed reference channel audio to get $M{-}1$ non-reference channel audio reconstructed.


\vspace{-5pt}
\section{Experiments and Dataset}
\vspace{-5pt}
\label{sec:Experiements}


We use the AISHELL-2 speech dataset~\cite{du2018aishell2} as clean speech sources to synthesize the multi-channel reverberant dataset. Our array configuration is from a real-world microphone array designed for the meeting scenario, which is an 8-channel linear non-uniform array where the distances between each pair of neighbor microphones are $[2,2,2,14,2,2,2]$ in centimeters. A total of 10k multi-channel RIRs are generated with random room characteristics using the image-source method. The reverberation time lies in the range [0, 0.7s]. A total of 90k, 7.5k, and 2k utterances are generated for the `Train', `Dev', and `Test' datasets. We use 16kHz sampling rate. 

\begin{table*}[hbt]
\vspace{-0.5cm}
\caption{Evaluation results for 8-channel reverberant audio reconstruction on the test set. CI denotes Channel Independent. SS denotes spatial similarity. SB-CODEC denotes our sub-band codec. DoA Error is in degrees while RTF Error is in radians. SIG, BAK, and OVRL are from DNSMOS \cite{reddy2022dnsmos} raw scores. MIMO E2E corresponds to our black-box end-to-end MIMO model as in Section~\ref{sec:results}.}
\vspace{-4pt}
\label{table:result}
\vspace{-5pt}
\begin{center}
\scriptsize
\begin{sc}
\renewcommand{\arraystretch}{1.1} 

\begin{tabular}{c|c|c|ccc|ccc|ccc} 
\hline
\multirow{3}{*}{\textbf{SYSTEM/METRICS}} & \multirow{3}{*}{\textbf{KBPS}} & \multirow{3}{*}{\textbf{CI}} & \multicolumn{6}{c|}{\textbf{BEAMFORMING}} & \multicolumn{3}{c}{\multirow{2}{*}{\textbf{SPATIAL}}}   \\
\cline{4-9}
& & & \multicolumn{3}{c|}{\textbf{Intrusive}}&\multicolumn{3}{c|}{\textbf{Non-Intrusive}}&&& \\
\cline{4-12}
& & & \textbf{SNR}$\mathbf{\uparrow}$ & \textbf{PESQ}$\mathbf{\uparrow}$ & \textbf{STOI}$\mathbf{\uparrow}$ & \textbf{SIG}$\mathbf{\uparrow}$ & 
\textbf{BAK}$\mathbf{\uparrow}$ &\textbf{OVRL}$\mathbf{\uparrow}$& \textbf{DoA Err}$\mathbf{\downarrow}$ & \textbf{RTF Err}$\mathbf{\downarrow}$ & \textbf{SS}$\mathbf{\uparrow}$ \\
\hline
\tiny\textbf{reverberant clean (groundtruth)} & \textbf{--} & \textbf{--} &$\infty$ & 4.5 & 1 &2.83 &3.50 & 2.38 &11.79 &0 &1 \\
\hline
\tiny\textbf{OPUS12} & 8*12=\textbf{96} & \multirow{5}{*}{\textbf{Yes}} & 7.13 & 2.53 & 0.79& 2.44 & 3.33 & 2.03 & 22.05 & 0.90 & 0.90\\

\cline{1-2}\cline{4-12}
\tiny\textbf{OPUS6} & \multirow{4}{*} {8*6=\textbf{48}} & & 0.91& 2.21 & 0.73 &2.12 & 2.16 & 1.77& 56.02 & 1.29& 0.85 \\
\cline{1-1}\cline{4-12}
\tiny\textbf{ENCODEC} \cite{encodec} & & &3.91 & 2.56 & 0.76 & 2.50&3.37 & 2.08& 56.88& 1.24&0.85\\
\cline{1-1}\cline{4-12}
\tiny\textbf{HIFI-CODEC} \cite{hificodec} & & &2.36 &2.62 & 0.77 &2.51& 3.39 & 2.09&56.88 &1.25 &0.86\\
\cline{1-1}\cline{4-12}
\tiny\textbf{SB-CODEC(Prop.)}  & & &\textbf{9.66} &2.71 & 0.79 & 2.57 &{3.30} & {2.15} &27.56 &0.86 &0.91 \\
\hline
\hline

\tiny\textbf{MIMO E2E(Prop.)} & \textbf{12} & \multirow{4}{*}{\textbf{No}} &6.14 &\textbf{3.02} & \textbf{0.85} &\textbf{2.99} & \textbf{3.80} & \textbf{2.56}&21.67 &0.95 &0.86\\
\cline{1-2}\cline{4-12}

\tiny\textbf{ENCODEC+SPATIALCODEC(Prop.)} & \multirow{3}{*}{6+6=\textbf{12}}&  &5.01 & 2.80 & 0.80 &2.73 &3.61 & 2.30&\textbf{13.78} &0.75 &\textbf{0.95}\\
\cline{1-1}\cline{4-12}
\tiny \textbf{HIFI-CODEC+SPATIALCODEC(Prop.)} & &  &4.32 & 2.87 & 0.82 &2.81 & 3.76 & 2.39&14.03 &0.75 &\textbf{0.95} \\
\cline{1-1}\cline{4-12}
\tiny \textbf{SB-CODEC+SPATIALCODEC(Prop.)} & &  & 8.28&2.92 & 0.82 & 2.86& 3.72 & 2.43&14.83 &\textbf{0.73} & \textbf{0.95}\\

\hline
\end{tabular}
\end{sc}
\end{center}
\vskip -0.3in
\vspace{-4pt}
\end{table*}

For our models, STFT is applied with 640-point hanning window size and 320-point hop size, resulting in 50 frames per second. For the complex ratio filter mentioned in Section~\ref{spatialcodec}, we use K=1 and L=4. For the vector quantization in the sub-band codec and SpatialCodec in Section~\ref{sec:Method}, all the codebooks have 1024 entries. We use 2 residual vector quantization layers so that both sub-band codec and SpatialCodec have 50*6*2*10~=~6kbps bitrate, 12kbps in total.

For training the reference channel codec, we follow the training scripts in HiFi-Codec \cite{hificodec}\footnote[1]{\url{https://github.com/yangdongchao/AcademiCodec}} and set the batch size to be 8 and the segment length to be 4 seconds. The training takes two million steps. The SpatialCodec is trained in a similar way. 
We use ADAM optimizer with $1\mathrm{e}{-4}$ learning rate.

\vspace{-5pt}
\section{Evaluation Metrics}
\vspace{-5pt}

This work focuses on spatial information preservation and reconstruction for multi-channel audio. For the reference channel codec, we use HiFi-Codec \cite{hificodec} and Encodec~\cite{encodec} as benchmark systems that can guarantee the performance of a single-channel audio codec. We mainly evaluate our methods using several spatial metrics and beamforming performance. We first use RTF (Relative Transfer Function) error~\cite{rtf}, and directional of arrival (DoA) estimation error with the classic MUSIC~\cite{MUSIC} algorithm. Then we propose a more intuitive spatial metric called spatial similarity. Besides the spatial metrics, we also evaluate the target DoA's beamforming performance to double-check both the signal quality and the main directivity of the target source. For all metrics in this section, we use 2048-point FFT, 512-point hop size, and 2048-point hanning window. Below are the detailed descriptions of each metric.

\vspace{-5pt}
\subsection{RTF Error}
\vspace{-5pt}
The RTF (Relative Transfer Function) error is defined in~\cite{rtf} as the angle between the groundtruth RTF $a(f)$ and estimated RTF $\hat{a}(f)$ averaged over all frequency bins:

\vspace{-5pt}
\begin{equation}
   \text{RTF\ Error} = \frac{1}{F} \sum \text{arccos\ Re}(\frac{\hat{a}(f)^\text{H}a(f)}{|\hat{a}(f)||a(f)|})
\end{equation}

where ground-truth RTF $a(f)$ and estimated RTF $\hat{a}(f)$ are extracted from the first principal component of the $M$-channel STFT matrix $X(f) \in \mathbb{C}^{M \times T}$ and $\hat{X}(f) \in \mathbb{C}^{M \times T}$, respectively.

\vspace{-5pt}
\subsection{MUSIC DoA Error}
\vspace{-5pt}
MUSIC DoA error is the DoA estimation error corresponding to the ground-truth DoA. This is a very intuitive metric to roughly measure the ability of the algorithm to capture the direct path component. We use the classic MUSIC~\cite{MUSIC} algorithm's implementation in pyroomacoustics~\cite{pyroomacoutstics}.

\vspace{-8pt}
\subsection{Spatial Similarity}
\vspace{-5pt}
We provide a spatially more intuitive metric which we call spatial similarity (SS). We first define a $B$ dimensional spatial feature vector $\textbf{S}(f)$, which are just $B$ fixed super-directive beamforming responses' magnitude averaged over time. Given a multi-channel input $X$, the spatial feature is defined as:

\vspace{-16pt}
\begin{align}\label{eq8}
\vspace{-14pt}
    Y_k (f) &= \frac{1}{T} \sum_t |\text{Beamformer}_b (X(t, f))|, b\in[1,...,B]\\
    \boldsymbol{S}(f) &= [Y_1(f),Y_2(f),Y_3(f), ..., Y_B(f)]
\end{align}
The spatial feature is similar to the beamspace-domain concept in~\cite{beamspace, LI2024101976}, which uses a few spatially sampled beamformers to transform the multi-channel STFT from RTF domain to the beamspace domain, which matches our spatial intuition. We uniformly sample beamformers' directions in the inter-channel time difference domain, which means all pairs of neighboring beamform directions have the same inter-channel time difference from the perspective of the linear array. The $b$-th beamformer's direction $\theta_{b}$ follows:

\vspace{-9pt}
\begin{equation}
\vspace{-3pt}
    \theta_{b} = \text{arccos} (1-\frac{2b}{B})
\end{equation}

For the super-directive beamforming, we set diagonal loading to be 1e-2. Details can be checked here~\cite{beamspace} or our \href{https://github.com/XZWY/SpatialCodec}{code}.

Then we define the spatial similarity based on the defined spatial features. After we get the spatial features of the original and reconstructed multi-channel audio, $\boldsymbol{S}(f)$ and $\boldsymbol{\hat{S}}(f)$ accordingly, the spatial similarity is defined as:

\vspace{-8pt}
\begin{equation}
    \text{Spatial\ Similarity} = \frac{1}{F}\sum_{f}\frac{\boldsymbol{S}(f)^T \boldsymbol{\hat{S}}(f)}{||\boldsymbol{S}(f)| |\cdot||\boldsymbol{\hat{S}}(f)||}
    \vspace{-5pt}
\end{equation}


We believe this spatial similarity is a more intuitive metric to measure spatial cue preservation for all directions. In our case, it can capture the direct path and early reflections' direction of arrival. We also set B=50 when calculating the spatial similarity.

\vspace{-5pt}
\subsection{Beamforming Performance}
\vspace{-5pt}
Besides beamforming towards a few fixed directions, we can also beamform towards the ground truth direct path's direction and check the beamforming result. This also provides a perceptual listening opportunity to check if the direct path's signal is preserved well or distorted in the source direction. We report PESQ, SNR, STOI, and the non-intrusive DNSMOS~\cite{reddy2022dnsmos} raw scores (SIG, BAK, OVRL).

\vspace{-5pt}
\section{Results and Discussions}
\label{sec:results}
\vspace{-5pt}

\begin{figure}[h]

  \centering
  \includegraphics[width=\linewidth]{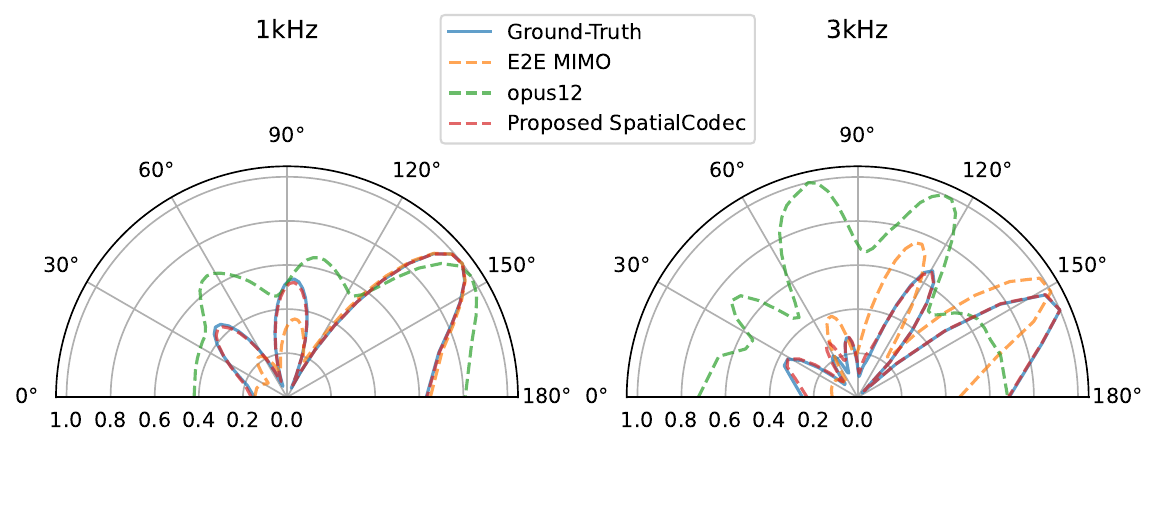}
  \vskip -0.35in

  \caption{Spatial Features Visualization (1kHz and 3kHz).}
  \label{fig:spatial_feature}
\vskip -0.1in
\vspace{-7pt}
\end{figure}

Our baselines include a few channel-independent coding methods (Use single channel codec to code each channel independently). They are OPUS \cite{opus} with two versions (6 and 12 kbps per channel), HiFi-Codec \cite{hificodec}, Encodec \cite{encodec}, and our proposed sub-band codec (all 6 kbps per channel). HiFi-Codec and Encodec are retrained on our dataset. We also design a black-box multi-channel in multi-channel out end-to-end (MIMO E2E) model for comparison and analysis, which has the same architecture as the SpatialCodec except the last decoder layer directly outputs the multi-channel STFT reconstruction. The number of residual vector quantization layers is set to 4 so the bitrate is 12kbps. 
For the SpatialCodec, We also substitute the first branch's sub-band codec with HiFi-Codec and Encodec as two other baselines. Note that HiFi-Codec and Encodec are time-domain models, so STFT is needed before applying the complex ratio filter.\\
\noindent \textbf{Overall Comparisons}: Table~\ref{table:result} shows the evaluation results of all proposed and baseline models. Generally, we can observe that our proposed SpatialCodec and MIMO E2E model achieves much better performance in all metrics than any other higher bitrate baselines, while maintaining only 12 kbps of bitrate in total. Also, we can see that channel-independent coding baselines are not able to preserve spatial information well even though each channel is coded with decent coding rates. Last, the black-box MIMO E2E model is worse than our proposed two-branch approach in spatial performance, e.g., spatial similarity (SS) 0.86 vs. 0.95, which shows it's hard to directly learn spatial preservation through such a black-box network. However, it performs better than SpatialCodec in beamforming performance because the main lobe of the beamformer is generally wide and not sharply focused on the target direction.\\
\noindent \textbf{Spatial performance}: For \textbf{DoA error}, SpatialCodec achieves the best result when combined with Enodec, which differs from the DoA error of the ground-truth reverberant clean signal by only 2 degrees. When combined with sub-band codec or Hifi-Codec, SpatialCodec still achieves similar performance. All channel-independent baselines have over 22 degrees of DoA error, which is much worse than our proposed SpatialCodec methods. For \textbf{RTF error}, our SpatialCodec has the best performance when combined with our sub-band codec. Overall, it achieves an RTF error of 0.73, which exceeds the OPUS12 (8*12=96 kbps) baseline by over 18\% and exceeds the OPUS6 (8*6=48 kbps) baseline by over 43\%. Again, channel-independent methods all have RTF errors higher than 0.8, showing a comparably large error. The black-box MIMO E2E model can only achieve 0.95 error which is even worse than OPUS12. For \textbf{spatial similarity (SS)}, our SpatialCodec achieves a superior score of 0.95 when combined with any reference channel neural codec. For baseline methods, only OPUS12 and sub-band codec have a spatial similarity of more than 0.9. The black-box MIMO E2E model again shows inferior results here. Figure~\ref{fig:spatial_feature} shows a visualization of B=50 normalized spatial features of four methods: original multi-channel audio, OPUS with two bitrate versions, and our proposed SpatialCodec (with sub-band codec). We can see that our SpatialCodec very closely aligns with the ground-truth feature in both 1 and 3kHz frequencies, while OPUS12 and MIMO E2E show quite some deviations from the ground-truth pattern. Overall, for spatial performance, the proposed SpatialCodec method exceeds other baselines when it's combined with any reference channel codecs. Note that both the black-box MIMO E2E model and high bitrate channel-independent coding methods are unable to achieve similar performance with our two-branch methods.\\
\noindent \textbf{Beamforming Performance}: Our SpatialCodec and MIMO E2E all have promising performance in both intrusive and non-intrusive beamforming metrics, showing decent spectral reconstruction. The reconstructed multi-channel audio demos are also given on our \href{https://xzwy.github.io/SpatialCodecDemo/}{website}. Overall, MIMO E2E performs the best performance, probably because it incorporates all channels to code spectral information. Then the SpatialCodec performs slightly worse but still much better than all other baselines. For \textbf{PESQ}, MIMO E2E has a score of 3.02, while the best SpatialCodec can achieve 2.92 (coupled with sub-band codec). OPUS12 can only achieve a score of 2.53 despite its high bitrate. Other neural channel-independent baselines achieve scores around 2.65, while sub-band codec performs a bit better. The situation is similar for \textbf{STOI}, with MIMO E2E scoring 0.85 and SpatialCodec scoring 0.8+, all other baselines score below 0.8. For \textbf{SNR}, the sub-band codec achieves the highest 9.66 dB, while our best SpatialCodec can achieve 8.28 dB. Then comes OPUS12's 7.13 dB and MIMO E2E's 6.14 dB. SNR is probably not the best metric here to show perceptual spectral performance but it's a good sanity check. For non-intrusive \textbf{DNSMOS} raw scores, the overall score is a bit low for all systems because the target signal is reverberant. MIMO E2E again performs the best. Together with MIMO E2E, SpatialCodec + sub-band Codec exceeds the performance of the ground-truth reverberant clean, because the original AISHELL2 \cite{du2018aishell2} speech data is not pure clean.
However, this shows our model has promising spectral performance. Furthermore, SpatialCodec performs significantly better than channel-independent methods. 

\vspace{-10pt}


\vspace{-7pt}
\section{Conclusion and Future Work}
\vspace{-10pt}
This paper proposes a novel two-branch codec framework for neural spatial speech coding. The first branch is a reference channel codec to code the source spectral information. The second branch aims to extract and code the spatial information for multi-channel resynthesis. We also propose a few novel metrics to measure spatial and spectral quality, including spatial similarity and beamforming performance. Our 12 kbps approach performs much better than all baselines including 96 kbps OPUS12. We also designed a black-box MIMO E2E model for comparison. Although we only test our system on single-speaker reverberant speech signals, this approach should be able to generalize to more complicated scenarios like multi-speaker, music sources, and moving sources. We leave these more challenging directions for future research.


\newpage

\bibliographystyle{IEEEbib}
\bibliography{main}

\end{document}